\documentstyle[12pt]{article}

\begin{document}

\font\ninerm = cmr9

\baselineskip 14pt plus .5pt minus .5pt

%\pageno=0\footline={\ifnum\pageno>0 \hss --\folio-- \hss \else\fi}
\def\footnoterule{\kern-3pt \hrule width \hsize \kern2.6pt}

\hsize=6.0truein
\vsize=9.0truein
\textheight 8.5truein
\textwidth 5.5truein
\voffset=-.4in
\hoffset=-.4in

%\preprint{}
\title{Vortex Solutions of Four-fermion Theory coupled to a 
Yang-Mills-Chern-Simons Gauge Field}
\author{Hyuk-jae Lee\thanks{e-mail : hjlee@theory.yonsei.ac.kr},
Joo Youl Lee\thanks{e-mail : jylee@phya.yonsei.ac.kr}  and Jae Hyung Yee\thanks{e-mail : jhyee@phya.yonsei.ac.kr} \\
Department of Physics and Natural Science Research Institute \\
Yonsei University,
Seoul, 151-742, Korea}

\maketitle
\begin{abstract}
We have constructed a four-fermion theory coupled to a Yang-Mills-Chern-Simons gauge field
which admits static multi-vortex solutions. This is achieved
through the introduction of an anomalous magnetic interation term in addition to the 
usual minimal coupling, and the appropriate choice of the fermion quartic coupling constant.
\end{abstract}

\baselineskip=16pt plus 3pt minus 3pt

\pagestyle{plain}
\pagenumbering{arabic}
\setcounter{page}{1}

\pagebreak[3]
\setcounter{equation}{0}
%\renewcommand{\theequation}{\arabic{section}.\arabic{equation}}
%\section{Introduction}
%\nopagebreak
%\medskip
%\nopagebreak
\newpage

Since the introduction of the Chern-Simons
action \cite{Tem} as a new possible gauge field 
theory in $(2+1)$-dimensional space-time, it has been successfully
applied to explain various $(2+1)$-dimensional 
phenomena including the high $T_c$ superconductivity and the
integral and fractional quantum Hall effects. 
The Chern-Simons term has also made it possible to construct various 
field theoretic models which possess classical vortex solutions with various physically
interesting properties \cite{Jackiw}-\cite{Nem}. They include the relativistic \cite{Jackiw} and
non-relativistic \cite{Pi} scalar field theories 
interacting with Abelian Chern-Simons fields, which admit static 
multi-vortex solutions saturating the Bogomol'nyi bound \cite{Bog} that reduces 
the second-order field equations to the first-order ones. The vortex solutions have
also been found for the scalar theories coupled to both the Maxwell and Chern-Simons terms \cite{Minn}. 
Such theories with static votex solutions have also been extended by introducing supersymmetry \cite{Min} 
and by adding a new interaction term such as anomalous magnetic interation term \cite{Antil}.

The Chern-Simons gauge theories coupled to the relativistic \cite{Li} and non-relativistic \cite{Duval} \cite{Nem}
fermion matter fields have also been found to admit static votex solutions. Recently it has been found 
that the four-fermion theory coupled to a Maxwell-Chern-Simons field admits static vortex solutions
that have an interesting physical property \cite{Lee}. In this theory two matter currents, 
the electromagnetic current and a new topological current associated with the electromagnetic 
current, couple to the gauge field. 
This may provide an interesting model for studying the dynamical properties of magnetic vortices from the field
theoretic point of view. It is the purpose of this paper to study the non-Abelian generalization of this model.

We consider the self-interacting spinor field theory coupled to a non-Abelian Yang-Mills-Chern-Simons
field described by the Lagrangian,
\begin{eqnarray}
{\cal L}&=&-\frac{1}{2}Tr({F}^{\mu \nu}{F}_{\mu \nu})+\frac{\kappa}{2}
\epsilon^{\mu \nu \rho} Tr({ F}_{\mu \nu}{ A}_{\rho} -\frac{1}{3}
{ A}_{\mu}[{ A}_{\nu}, { A}_{\rho}])\nonumber\\
& &+i\bar\psi\gamma^{\mu}
 {\cal D}_{\mu}\psi -m\bar\psi \psi +\frac{1}{2} g (\bar\psi { T}^a \psi)(\bar\psi { T}^a \psi),
\label{La}
\end{eqnarray}
where  $\gamma$ matrices are chosen to be
\begin{equation}
\gamma^0 = \sigma^3 ,\,\,\, \gamma^1 = i\sigma^1 ,\,\,\,
 \gamma^2 =i\sigma^2,
\label{Gama}
\end{equation}
in terms of the Pauli matrices $\sigma^i$, the vector potentials are represented as  anti-Hermitian matrices as
\begin{equation}
{ A}_{\mu}=A_{\mu}^a { T}^a,
\label{Au}
\end{equation}
with the group generators $T^a$ satisfying
\begin{equation}
[{ T}^a, { T}^b]=f^{abc}{ T}^c
\label{Tu}
\end{equation}
\begin{equation}
({ T}^a)^{\dagger}=-{ T}^a,
\label{T}
\end{equation}
and the spinor field $\psi$ transforms as an irreducible representation of the Lie group generated by $T^a$.
We introduce the covariant derivative ${\cal D}_{\mu}$ including both 
the usual minimal coupling and the magnetic moment interaction with
the magnetic moment $u$ \cite{Antil},
\begin{equation}
{\cal D}_{\mu}\psi = D_{\mu}\psi +\frac{u}{4}
\epsilon_{\mu \nu \rho} { F}^{\nu \rho} \psi
\label{Du}
\end{equation}
where $D_{\mu}= \partial_{\mu} +{ A}_{\mu}$.

The equations of motion are
\begin{equation}
 i\gamma^{\mu} {\cal D}_{\mu} \psi -m \psi +g{ T}^a(\bar \psi { T}^a \psi)\psi =0,
\label{maf}
\end{equation}
\begin{equation}
D_{\alpha}{ F}^{\alpha \beta ,a} +\frac{\kappa}{2} \epsilon^{\beta
\alpha \lambda}{ F}_{\alpha \lambda} ^a = -i\bar\psi\gamma^{\beta}{ T}^a
\psi -i\frac{u}{2} \epsilon^{\beta \alpha \lambda} D_{\alpha}(\bar
\psi\gamma_{\lambda} { T}^a \psi).
\label{csf}
\end{equation}
Note that the anti-Hermitian matrix version of
the current density reads
\begin{equation}
j_{\mu}={ T}^a j_{\mu} ^a =-i{ T}^a(\bar \psi \gamma_{\mu} { T}^a \psi),
\end{equation}
and  the matter density is defined as
\begin{equation}
\rho={ T}^a \rho^a=-i{ T}^a(\psi^\dagger { T}^a\psi).
\label{rho}
\end{equation}

We will show that the system described by the Lagrangian (\ref{La}) supports static vortex solutions.
For this purpose, we choose the temporal gauge, ${A}_0 =0$, and consider the gauge field ${A}_i$ to be
static.
We take the fermion field $\psi$ in component form,
 $\psi={ \psi_+ \choose \psi_-}e^{-iE_f t}$, where $E_f$ is a constant, and ${1 \choose 0}$ is the spin-up and
${0 \choose 1}$ the spin-down Pauli spinors \cite{Li,Lee}.
The equation of motion (\ref{maf}) can then be written as coupled equations for
$\psi_+$ and $\psi_-$.
If we choose the spinor field as
\begin{equation}
\psi=\psi_+{1 \choose 0}e^{-iE_f t},
\label{Psi1}
\end{equation}
then the field equation (\ref{maf}) reduces to
\begin{equation}
 (E_f -m +ig\rho_+ ^a { T}^a +i\frac{u}{2}F^{1 2 ,a}{ T}^a)\psi_+=0,
\label{eq1}
\end{equation}
\begin{equation}
{\cal D}_+\psi_+  =0,
\label{sdeq1}
\end{equation}
due to the choise of $\gamma$ matrices (\ref{Gama}), and if we take
\begin{equation}
\psi=\psi_-{0 \choose 1}e^{-iE_f t},
\label{Psi2}
\end{equation}
Eq.(\ref{maf}) reduces to
\begin{equation}
 (-E_f -m -ig\rho_- ^a{ T}^a -i\frac{u}{2}F^{1 2 ,a}{ T}^a)\psi_-=0,
\label{eq2}
\end{equation}
\begin{equation}
{\cal D}_-\psi_- =0,
\label{sdeq2}
\end{equation}
where ${\cal D}_\pm ={\cal D}_1 \pm i{\cal D}_2$  and
${\rho}_{\pm} =-iT^a(\psi_{\pm}^{\dagger} T^a \psi_{\pm})$.
${\cal D}_i =D_i$, for \,$i=1,2$ \,\, since we have chosen the gauge condition, $A_0 =0$.
Above equations show that the fermion fields, $\psi_+$ and $\psi_-$,
satisfy the self-dual equations.

From Eqs.(\ref{Gama}), (\ref{Psi1}) and (\ref{Psi2}), we find 
${\bar {\psi}}{\gamma}^i {\psi}=0$ , which implies that
${\epsilon^{0ij}}D_i ({\bar {\psi}}\gamma_j \psi)=0$, for $i, j =1,2$.
Then due to Eqs.(\ref{Psi1}) and (\ref{Psi2}),  Eq.(\ref{csf}) reduces to
\begin{equation}
F_{12}^a { T}^a \equiv -{ B} =\frac{1}{\kappa}\rho_{\pm} ^a { T}^a
\label{Bi}
\end{equation}
\begin{equation}
D_{i}F^{ij ,a}=-\frac{u}{2}\epsilon^{0ij}D_{i}\rho_{\pm}^a,
\label{Fi}
\end{equation}
where (\ref{Bi}) is the Gauss' law constraint.
For the above two equations to be consistent constants $u$ and $\kappa$ must satisfy the condition
\begin{equation}
u= -\frac{2}{\kappa}.
\label{bbb}
\end{equation}
This shows that for the theory (\ref{La}) to have consistent static field equations, one need to introduce the 
anomalous magnetic interation term.
For the Eqs.(\ref{eq1}) and (\ref{eq2}) to be consistent with the Gauss' law constraint (\ref{Bi}), 
the quartic coupling constant must satisfy,
\begin{equation}
g=\frac{u^2}{4},
\label{gu}
\end{equation}
and
\begin{eqnarray}
(E_f -m)\psi_+ &=&0,\nonumber\\
(-E_f -m)\psi_- &=&0,
\label{qqq}
\end{eqnarray}
which determine the constant $E_f$ for the solutions (\ref{Psi1}) and (\ref{Psi2}), respectively.
The self-dual equations (\ref{sdeq1}) and (\ref{sdeq2}) then become
\begin{equation}
D_{\pm} {\psi}_{\pm} =0.
\label{dual1}
\end{equation}
This result corresponds to the fermion version of the non-Abelian generalization \cite{Dun} 
\cite{JaPi} of the Jackiw-Pi model \cite{Pi}.

To solve the self-dual equations,
we use the adjoint representation for the matter fields
and define the matter field matrix ${ {\Psi}}$ by contracting the multiplet ${\psi}$ with
generators of the Lie algebra. We usually denote the representation of the generators by
${\cal T}^a=T^a$ (${\cal T}^a$ is $(2i)^{-1}$ times the Pauli matrices or Gell-Mann matrices
for $SU(2)$ and $SU(3)$, respectively):
\begin{equation}
{ \Psi}_{mn}=\psi_a({\cal T}^a)_{mn}.
\label{Psimn}
\end{equation}
In this representation,
the self-dual matter field equation, for the ${\psi}_+$ field of Eq. (\ref{Psi1}), may be written as
\begin{equation}
\partial_+{ \Psi} +[A_+ , { \Psi}] =0
\label{adeqp1}
\end{equation}
\begin{equation}
\partial_-{ \Psi^\dagger} +[A_- , { \Psi^\dagger}] =0,
\label{adeqn1}
\end{equation}
where $\partial_{\pm}=\partial_1 \pm i\partial_2$ and $A_{\pm} =A_1 \pm i A_2$.
The matter density (\ref{rho}) then reads
\begin{equation}
\rho=-i{\cal T}^a ({\psi}^{\dagger}_m f_{man}{\psi}_n)
=i\psi^{\dagger}_m[{\cal T}^m ,{\cal T}^n]\psi_n
=-i[\Psi^{\dagger} ,\Psi],
\end{equation}
and the Chern-Simons equation (\ref{Bi}) becomes
\begin{equation}
\partial_- A_+ -\partial_+ A_- +[A_-, A_+] =\frac{2i}{\kappa}\rho.
\label{adeqcs1}
\end{equation}
For the field ${\psi}_-$ of Eq. (\ref{Psi2}), the equations may be written as
\begin{equation}
\partial_-{ \Psi} +[A_- , { \Psi}] =0
\label{adeqn2}
\end{equation}
\begin{equation}
\partial_+{ \Psi^\dagger} +[A_+ , { \Psi^\dagger}] =0
\label{adeqp2}
\end{equation}
\begin{equation}
\partial_- A_+ -\partial_+ A_- +[A_-, A_+] =\frac{2i}{\kappa}\rho .
\label{adeqcs2}
\end{equation}

In order to find the solutions of these field equations we employ some standard Lie group notation \cite{Dun} \cite{JaPi}.
The group generators are given in the Cartan-Weyl basis, with the
commuting set which comprises the Cartan subalgebra, denoted by
$H^i =(H^i)^{\dagger}$, and the ladder generators denoted by $E^{\pm n}=(E^{\mp n})^{\dagger}$.
The index $i$ ranges over the rank $r$ of the group, while $n$ ranges up to $s$ such that
$2s+r=d$, the dimension of the group.
It is always possible to select an $r$-member subset of the ladder operators that
satisfy
\begin{equation}
[E^n , E^{-n'}] = \delta_{nn'} \sum_{i=1}^r v_n^i H^i,
\label{En}
\end{equation}
\begin{equation}
[H^i , E^{\pm n}] = {\pm}v_n ^i E^{\pm n},
\end{equation}
and the $v_n^i =-v_{-n}^i$ comprise $s$ real "root vectors" for $n=1\cdots s$ with $r$ components,
$i=1\cdots r$.
Introducing a new symbol, $e^{\alpha}=c_{\alpha}E^{\alpha}$ such that
$c_{\alpha}$ is a numerical factor, we have
\begin{equation}
[e^{\alpha}, e^{-\alpha '}]=\delta_{\alpha \alpha '}h^{\alpha}
\label{ea}
\end{equation}
\begin{equation}
[h^{\alpha}, e^{-\beta}]=K_{\beta \alpha}e^{\beta},
\label{eb}
\end{equation}
where
\begin{equation}
h^{\alpha}=|c_{\alpha}|^2 \sum_{i=1}^r v_{\alpha}^i H^i
\label{ca}
\end{equation}
\begin{equation}
K_{\alpha \beta}=|c_{\beta}|^2 \sum_{i=1}^r v_{\alpha}^i v_{\beta}^i.
\label{Kab}
\end{equation}
$K_{\alpha \beta}$ is called the Cartan matrix.
Now we will show that a special Ans\"atz reduces the self-dual equations to integrable nonlinear equations.
This is achieved by using the field decomposition in the form,
\begin{eqnarray}
{ \Psi} &=& \sum_{\alpha =1}^r u_{\alpha}e^{\alpha}\label{P} \\
A_-&=&\sum_{\alpha=1}^r A_{\alpha}h^{\alpha}\label{An} \\
A_+ &=& -\sum_{\alpha=1}^r A_{\alpha}^* h^{\alpha}\label{Ap}.
\end{eqnarray}

For the ${\psi}_+$ field, the self-dual matter field equation 
reads 
\begin{equation}
\partial_+ u_{\alpha}-u_{\alpha}\sum_{\beta=1}^r K_{\alpha \beta}A_{\beta}^* =0.
\label{up}
\end{equation}
This equation can be solved for $A_{\alpha}^*$:
\begin{equation}
A_{\alpha}^* =\sum_{\beta}^r K_{\alpha \beta}^{-1}\partial_+ \log u_{\beta}.
\label{Aap}
\end{equation}
Similarly, for the ${\psi}_-$ field, we find,
\begin{equation}
\partial_- u_{\alpha}+u_{\alpha} \sum_{\beta=1}^r K_{\alpha \beta}A_{\beta}-it (\partial_- E_f)u_{\alpha} =0,
\label{un}
\end{equation}
\begin{equation}
A_{\alpha}=-\sum_{\beta=1}^r K_{\alpha \beta}^{-1}\partial_- \log u_{\beta}.
\label{Aan}
\end{equation}

The Chern-Simons equations, Eqs (\ref{adeqcs1}) or (\ref{adeqcs2}), can then be written,
by virtue of (\ref{An}) and (\ref{Ap}), as
\begin{equation}
\partial_- A_{\alpha}^* + \partial_+ A_{\alpha} =\frac{2}{\kappa} \rho_{\alpha},
\label{cseq}
\end{equation}
where $\rho_{\alpha} \equiv u^{\dagger}_{\alpha} u_{\alpha}$ for this representation.
From Eq.(\ref{cseq}) and Eq.(\ref{Aap}) or (\ref{Aan}), we finally find that the matter 
density ${\rho}_{\alpha}$ satisfies the Toda equation,
\begin{equation}
\nabla^2 \ln \rho_{\alpha} =\pm \frac{2}{\kappa} \sum_{\beta =1}^r K_{\alpha \beta}\rho_{\beta},
\label{Toda}
\end{equation}
where the $\pm$ signs are for the solutions for the cases (\ref{Psi1}) and (\ref{Psi2}), respectively.
From this Toda equation,  we can find the solutions for the general gauge group $SU(N)$.
This shows that one can find the solutions for the self-dual equations of the theory (\ref{La}) 
for general group $SU(N)$.

The total energy of the system can be written as
\begin{eqnarray}
 E&=&\int d^2 r{\cal H}
=\int d^2 r[\frac{1}{2}(F^a_{12} +\frac{u}{2}\rho^a _\pm)(F^a_{12} +\frac{u}{2}\rho^a _\pm)
 +\frac{1}{2}(g -\frac{u^2}{4})\rho^a _\pm \rho^a _\pm \pm m\psi^{\dagger} _{\pm} \psi_{\pm}]\nonumber \\
  &=&\pm m\int d^2 r \psi^{\dagger}_\pm \psi_{\pm},
\label{energy}
\end{eqnarray}
where $\psi^{\dagger }_\pm \psi_\pm=\sum_i \psi^{\dagger i}_{\pm} \psi_{\pm}^i$ with $i$ denoting the
components of the field multiplet, and we have used the Gauss' law constraint and the consistency condition (\ref{bbb})
and (\ref{gu}). Note that these consistency conditions are such that all the quartic interaction terms in 
the Hamiltonian cancell out as in the Jackiw-Pi model.
 
The Toda equation (\ref{Toda}) for general $SU(N)$ is not soluble in closed form. 
For the case of $SU(2)$, however, Eq.(\ref{Toda}) reduces to the Liouville equation,
\begin{equation}
{\nabla}^2 ln{\rho}_{\pm}={\pm}{\frac{4}{\kappa}}{\rho}_{\pm},
\label{Lio}
\end{equation}
which is completely integrable.

If we take the case of Eq.(\ref{Psi1}), ${\rho}_{\alpha}$ becomes ${\rho}_+$ and ${\kappa}<0$
is required in order to have nonsingular positive charge density ${\rho}_+$.
If we take the case of Eq.(\ref{Psi2}), on the other hand, ${\rho}_{\alpha}={\rho}_-$ and ${\kappa}>0$
is required for the nonsingular density ${\rho}_-$.
That is, both solutions involve only one of the $(2+1)$-dimensional spinor field components,
depending on the sign of ${\kappa}$.
This corresponds to the embedding of $U(1)$ into $SU(2)$. 

The most general circularly symmetric nonsingular solutions to the Liouville equations (\ref{Lio}) 
involve
two positive constants $r_{\pm}$ and ${\cal N}_{\pm}$ \cite{Pi}:
\begin{equation}
{\rho}_{\pm} = {\mp} \frac{2{\kappa}{\cal N}_{\pm}^2}{r^2}[(\frac{r_{\mp}}{r})^{{\cal N}_{\pm}}
+(\frac{r}{r_{\mp}})^{{\cal N}_{\pm}}]^{-2},
\label{Sol}
\end{equation}
where $r_{\mp}$ are scale parameters and the $(-)$ sign for the negative $\kappa$ and
the $(+)$ sign for the positive $\kappa$.
To fix ${\cal N}_{\pm}$ , we observe that regularity at the origin,
${\rho}_{\pm} {\sim} r^{2{\cal N}_{\pm} -2}$ as ${r{\rightarrow}0}$, and at infinity,
${\rho}_{\pm} {\sim} r^{-2{\cal N}_{\pm}-2}$ as ${r{\rightarrow}{\infty}}$, requires
${\cal N}_{\pm}{\ge}1$. Especially for single-valuedness of ${\psi}_{\pm}$, ${\cal N}_{\pm}$
must be an integer \cite{Lee, JaPi}. The total charge of the soliton is then given by
\begin{equation}
Q_{\pm} = \int {\rho}_{\pm} d^2 r ={\mp}2{\pi}{\kappa}{\cal N}_{\pm} >0,
\end{equation}
which is the same as that of ref. \cite{Dun} \cite{JaPi}.
From the solution (\ref{Sol}), for the $SU(2)$ case, the total energy of the system
can be shown to be
\begin{equation}
E= {\pm}m(2{\pi}|{\kappa}|{\cal N}_{\pm}).
\label{engy}
\end{equation}

We have thus extended the four-fermion theory coupled to Maxwell-Chern-Simons field \cite{Lee}, 
which admits static multi-vortex solutions, to the one with non-Abelian symmetry.
We have shown that this non-Abelian model also admits the multi-vortex solutions. 
This is achieved through the introduction of an anomalous magnetic interation term in addition to
the conventional minimal gauge coupling, which is responsible for the right-hand side 
of Eq.(\ref{Fi}) that guarantees the consistent Gauss' law constraint, Eq.(\ref{Bi}).
Although the multi-vortex solutions for the $SU(2)$ case are the same as those of Jackiw-Pi model, the moduli space
dynamics \cite{manton} of these solutions will be quite different due to the Maxwell term in the 
Lagrangian which is quadratic in time derivatives of the gauge field.

\begin{center}
{\bf Acknowledgment}
\end{center}
This work was supported in part by Korea Science and Engineering Foundation under Grants Nos.
065-0200-001-2 and 97-07-02-02-01-3, by the Center for Theoretical Physics(SNU), and by the 
Basic Science Research Institute Program, Ministry of Education, under Project No. BSRI-97-2425.
H.-j. Lee wishes also to acknowledge to the financial support of Korea Research Foundation 
made in the program year of 1997.


\begin{thebibliography}{99}
\bibitem{Tem} R. Jackiw and S. Templeton, Phys. Rev. {\bf D23}, 2291(1981).\\
J. Schonfeld, Nucl. Phys. {\bf B185}, 157(1981).\\
R. Jackiw and S. Templeton, Phys. Rev. Lett. {\bf 48}, 975(1982).\\
S. Deser, R. Jackiw and S. Templeton, Ann. Phys. (N.Y.){\bf 140}, 372(1982).\\
C. R. Hagen, Ann. Phys. (N. Y.){\bf 157}, 342(1984).
\bibitem{Jackiw} J. Hong, Y. Kim and P. Y. Pac, Phys. Rev. Lett. {\bf 64} ,2230(1990)\\
R. Jackiw and E. J. Weinberg, Phys. Rev. Lett. {\bf 64} ,2234(1990).
\bibitem{Pi} R. Jackiw and S. Y. Pi, Phys. Rev. Lett. {\bf 64} ,2969(1990).
%\bibitem{Nielsen} H. B. Nielsen and P. Olesen, Nucl. Phys. {\bf B61} ,45(1973).
\bibitem{Bog}E. B. Bogomol'nyi, Sov. J. Nucl. Phys. {\bf 24}, 449(1976).
\bibitem{Minn}S. K. Paul and A. Khare, Phys. Lett. {\bf B174}, 420(1986).\\
C. Lee, K, Lee and H, Min, Phys. Lett. {\bf B252}, 79(1990).
\bibitem{Min} T. Lee and H. Min, Phys. Rev. {\bf D50} ,R7738(1994).
\bibitem{Antil} M. Torres, Phys. Rev. {\bf D46}(1992)R2295.\\
A. Antill\' on, J. Escalona, G. Germ\'an and M. Torres, Phys. Lett. {\bf B359} ,327(1995).
\bibitem{Li} S. Li and R. K. Bhaduri, Phys. Rev. {\bf D43} ,3573(1991).\\
J. Shin and J. H. Yee, Phys. Rev. {\bf D50}, 4223(1994).
\bibitem{Duval} C. Duval, P. A. Horv\'athy and L. Palla, Ann. Phys. {\bf 249}, 265(1995).
\bibitem{Nem} Z. N\'emeth, Phys. Rev. {\bf D56}, 5066(1997).
\bibitem{Lee} S. Hyun, J. Shin, J. H. Yee and H.-j. Lee, Phys. Rev. {\bf D55} ,3900(1997).
\bibitem{Dun} B. Grossman, Phys. Rev. Lett. {\bf 65}, 3230(1990).\\
G. V. Dunne, R. Jackiw, S. Y. Pi, and C. A. Trugenberger, Phys.Rev. {\bf D43} ,1332(1991).
\bibitem{JaPi} R. Jackiw and S. Y. Pi, Prog. Theor. Phys. Suppl. {\bf 107}, 1(1992).
\bibitem{manton}N. S. Manton, Phys. Lett. {\bf B110}, 54(1982); ibid {\bf B154}, 397(1985).
\end{thebibliography}
\end{document}